\documentclass{article}

\usepackage{latexsym}

\newcommand{\SB}{\{\,}
\newcommand{\SM}{\;{:}\;}
\newcommand{\SE}{\,\}}
\newcommand{\Card}[1]{|#1|}
\newcommand{\hy}{\hbox{-}\nobreak\hskip0pt}
\newcommand{\prf}{\mbox{\normalfont prf}}
\newcommand{\OR}{\mbox{\normalfont OR}}

\newcommand{\NP}{\mbox{\normalfont NP}}
\newenvironment{proof}{\pf}{\qed}
\newcommand{\swap}{\mbox{\normalfont swap}}

\begin{document}
\newcommand{\2}{\vspace{0.2 cm}}
\newcommand{\dom}{\mbox{$\rightarrow$}}
\newcommand{\ndom}{\mbox{$\not\rightarrow$}}
\newcommand{\sdom}{\mbox{$\Rightarrow$}}
\newcommand{\nsdom}{\mbox{$\not\Rightarrow$}}
\newcommand{\qed}{\hfill$\Box$}
\newcommand{\pf}{{\bf Proof: }}
\newtheorem{theorem}{Theorem}[section]
\newtheorem{algorithm}[theorem]{Algorithm}
\newtheorem{proposition}[theorem]{Proposition}
\newtheorem{lemma}[theorem]{Lemma}
\newtheorem{problem}[theorem]{Problem}
\newtheorem{corollary}[theorem]{Corollary}
\newtheorem{conjecture}[theorem]{Conjecture}
\newtheorem{krule}[theorem]{Kernelization Rule}
\newtheorem{remark}[theorem]{Remark}
\newcommand{\beq}{\begin{equation}}
\newcommand{\eeq}{\end{equation}}
\newcommand{\ra}{\rangle}
\newcommand{\la}{\langle}
\newcommand{\har}{\rightleftharpoons}
\newcommand{\<}[1]{\mbox{$\la #1 \ra$}}

\title{Fixed-Parameter Complexity of Minimum Profile Problems}

\date{}

\author
{Gregory Gutin\thanks{Corresponding author. Department of Computer
Science, Royal Holloway University of London, Egham, Surrey TW20
0EX, UK, gutin@cs.rhul.ac.uk and Department of Computer Science,
University of Haifa, Israel} \and  Stefan Szeider\thanks{ Department
of Computer Science, Durham University Science Labs, South Road,
Durham DH1 3LE, UK, stefan.szeider@durham.ac.uk} \and Anders
Yeo\thanks{Department of Computer Science, Royal Holloway University
of London, Egham, Surrey TW20 0EX, UK, anders@cs.rhul.ac.uk}}

\maketitle

\begin{abstract}
Let $G=(V,E)$ be a graph. An ordering of $G$ is a bijection
$\alpha:\ V\dom \{1,2,\ldots, |V|\}.$ For a vertex $v$ in $G$, its
closed neighborhood is $N[v]=\{u\in V:\ uv\in E\}\cup \{v\}.$ The
profile of an ordering $\alpha$ of $G$ is
$\prf_{\alpha}(G)=\sum_{v\in V}(\alpha(v)-\min\{\alpha(u):\ u\in
N[v]\}).$ The profile $\prf(G)$ of $G$ is the minimum of
$\prf_{\alpha}(G)$ over all orderings $\alpha$ of $G$. It is
well-known that $\prf(G)$ is the minimum number of edges in an
interval graph $H$ that contains $G$ is a subgraph. Since $|V|-1$ is
a tight lower bound for the profile of connected graphs $G=(V,E)$,
the parametrization above the guaranteed value $|V|-1$ is of
particular interest.  We show that deciding whether the profile of a
connected graph $G=(V,E)$ is at most $|V|-1+k$ is fixed-parameter
tractable with respect to the parameter $k$. We achieve this result
by reduction to a problem kernel of linear size.
\end{abstract}

\section{Introduction}

A parameterized problem $\Pi$ can be considered as a set of pairs
$(I,k)$ where $I$ is the \emph{problem instance} and $k$ (usually an
integer) is the \emph{parameter}. $\Pi$ is called
\emph{fixed-parameter tractable (FPT)} if membership of $(I,k)$ in
$\Pi$ can be decided in time $O(f(k)|I|^c)$, where $|I|$ is the size
of $I$, $f(k)$ is a computable function, and $c$ is a constant
independent from $k$ and $I$.

Let $G=(V,E)$ be a graph. An {\em ordering} of $G$ is a bijection
$\alpha:\ V\dom \{1,2,\ldots, |V|\}.$ We denote the set of orderings
of $G$ by $\OR(G)$. For a vertex $v$ in $G$, its {\em neighborhood}
is $N(v)=\{u\in V:\ uv\in E\}$ and its {\em closed neighborhood} is
$N[v]=N(v)\cup \{v\}.$ The {\em profile of a vertex} $z$ of $G$ in
an ordering $\alpha$ of $G$ is $\prf_{\alpha}(G,z) =\alpha(z) -
\min\{ \alpha(w) \mbox{ : } w \in N[z] \}).$ The {\em profile of a
set $Z\subseteq V$ in an ordering $\alpha$ of $G$} is
$\prf_{\alpha}(G,Z) = \sum_{z \in Z} \prf_{\alpha}(G,z).$ The {\em
profile of an ordering $\alpha$ of $G$} is
$\prf_{\alpha}(G)=\prf_{\alpha}(G,V).$ An ordering $\alpha$ of $G$
is {\em optimal} if $\prf_{\alpha}(G)=\min\{\prf_{\beta}(G):\
\beta\in \OR(G)\}.$ If $\alpha$ is optimal, then
$\prf(G)=\prf_{\alpha}(G)$ is called the {\em profile} of $G.$

In \cite{fominSIAMJDM13} Fomin and Golovach established equivalence
of $\prf(G)$ to other parameters including one important in graph
searching. Further areas of application of the profile and
equivalent parameters include computational biology
\cite{bodlCAB11,goldbergJCB2}, archaeology \cite{kendallPJM28} and
clone fingerprinting \cite{karpSTOC25}. The following is a
well-known NP-complete problem \cite{diazLNCS519,linAMAS10}.
 \vspace{-0.3cm}
\begin{quote}
  \noindent{\bfseries Minimum Profile Problem} (MPP)\\
  \emph{Instance:} A graph $G=(V,E)$ and a positive integer $k$.\\
  \emph{Question:} Does $G$ have an ordering of profile $\le k$?
\end{quote}
\vspace{-0.3cm}
 In fact this problem is equivalent to the following
problem that have been proved to be NP-complete even earlier (see
\cite{garey1979}). A graph $G=(V,E)$ is {\em interval} if we can
associate each vertex $v\in V$ with a closed interval $I_v$ in the
real line such that two distinct vertices $x$ and $y$ are adjacent
in $G$ if and only if $I_x\cap I_y\neq \emptyset.$
 \vspace{-0.3cm}
\begin{quote}
  \noindent{\bfseries Interval Graph Completion} (IGC)\\
  \emph{Instance:} A graph $G=(V,E)$ and a positive integer $k\ge |E|$.\\
  \emph{Question:} Is there a supergraph $G'$ of $G$ such that $G'$
  is an interval graph and it contains at most $k$ edges?
\end{quote}
\vspace{-0.3cm}
 The equivalence between MPP and IGC follows from the
next result:
\begin{theorem}\cite{billionnetRAIRO20}\label{bilth}
For any graph $G$ $\prf(G)$ equals the smallest number of edges in
an interval supergraph of $G$.
\end{theorem}

Thus, for every graph $G$, $\prf(E(G))\ge |E|.$ Hence, the following
parameterized problem is FPT.
 \vspace{-0.3cm}
\begin{quote}
  \noindent{\bfseries Profile Problem} (PP)\\
  \emph{Instance:} A graph $G=(V,E).$
  \emph{Parameter:} A positive integer $k$.\\
  \emph{Question:} Does $G$ have an ordering of profile $\le k$?
\end{quote}
 \vspace{-0.3cm}
Several authors consider the following  much more interesting and
problem; in fact, it is unknown whether the problem is FPT (private
communications with L. Cai, F. Fomin and H. Kaplan).
 \vspace{-0.3cm}
\begin{quote}
  \noindent{\bfseries Profile Above Guaranteed Value} (PAGV)\\
  \emph{Instance:} A graph $G=(V,E).$
  \emph{Parameter:} A positive integer $k$.\\
  \emph{Question:} Does $G$ have an ordering of profile $\le |E|+k$?
\end{quote}
 \vspace{-0.3cm}
 Unfortunately, we are not able to determine the
complexity of this problem. In this paper, we consider a somewhat
weaker version of MPAGV. We restrict ourselves to connected graphs
(the case of general graphs can be reduced to connected graphs).
Since $|E|\ge |V|-1$ for a connected graph $G=(V,E),$ consider the
following:
 \vspace{-0.3cm}
\begin{quote}
  \noindent{\bfseries Profile Above Vertex Guaranteed Value} (PAVGV)\\
  \emph{Instance:} A connected graph $G=(V,E).$
  \emph{Parameter:} A positive integer $k$.\\
  \emph{Question:} Does $G$ have an ordering of profile $\le |V|-1+k$?
\end{quote}
 \vspace{-0.3cm}
 This problem is of interest also because of Problem
VAP by Serna and Thilikos \cite{sernaEATCSB86} (see Section
\ref{npsec}). We will prove by means of a kernelization scheme that
the problem PAVGV is fixed-parameter tractable.

\section{Preliminary Results}

Let $G=(V,E)$ be a graph. If $X \subseteq V$ and $\alpha$ is an
ordering of $G$, then let $\alpha_X$ denote the ordering of $G-X$
in which $\alpha_X(u)<\alpha_X(v)$ if and only if
$\alpha(u)<\alpha(v)$ for all $u,v \in V(G)-X$. If $X=\{x\}$, then
we simply write $\alpha_x$ instead of $\alpha_{\{x\}}$.

\begin{lemma} \label{G_X}
Let $G=(V,E)$ be a graph of order $n$ and let $X$ be a set of
vertices such that $G-X$ is connected. If an optimal ordering
$\alpha$ has $\{\alpha^{-1}(1),\alpha^{-1}(n)\} \subseteq V(G-X)$
then $\prf_{\alpha}(G,V-X) \geq \prf_{\alpha_X}(G-X) + |X|$.
\end{lemma}
\begin{proof} Let $X=\{x_1,x_2,\ldots,x_r\}$ and define
$X_i=\{x_1,x_2,\ldots,x_i\}$ for all $0 \leq i \leq r$. We will by
induction show the following: (*) $\prf_{\alpha_{X_i}}(G-X_i,V-X)
\geq \prf_{\alpha_X}(G-X) + |X|-i.$ The above is clearly true when
$i=r$ as $X_r=X$ and $|X|=r$. If we can show that (*) is true for
$i=0$, then we are done. We will assume that (*) is true for some
$i>0$.

Since $G-X$ is connected and $\{\alpha^{-1}(1),\alpha^{-1}(n)\}
\subseteq V(G-X)$, there is an edge $uv \in E(G-X)$ such that
$\alpha_{X_{i-1}}(u) > \alpha_{X_{i-1}} (x_i) >
\alpha_{X_{i-1}}(v)$. This implies that the profile of $u$ is one
larger in $\alpha_{X_{i-1}}$ than it is in $\alpha_{X_i}$. This
implies the following:
$$\prf_{\alpha_{X_{i-1}}}(G-X_{i-1},G-X) \geq  \prf_{\alpha_{X_i}}(G-X_i,G-X)+1 \geq
\prf_{\alpha_X}(G-X) + |X|-i+1.$$ We are now done by induction.
\end{proof}

\begin{lemma} \label{prf_cycle} \cite{linAMAS10}
(i) If $G$ is a connected graph with $n$ vertices, then $\prf(G)
\geq n-1.$

(ii) Let $C_n$ denote a cycle with $n$ vertices. Then $\prf(C_n) =
2n-3$.
\end{lemma}

For a vertex $x$, $d(x)$ denotes its degree, i.e.,  $d(x)=|N(x)|.$ A
slightly weaker version of the following lemma is stated in
\cite{linAMAS10} without a proof.

\begin{lemma} \label{add_vertex}
If $G$ is an arbitrary graph of order $n$, $x \in V(G)$ and $\alpha$
is an optimal ordering of $G$, then $\prf_{\alpha}(G) \geq
\prf_{\alpha_x}(G-x)+d(x)$.
\end{lemma}
\begin{proof} Let $\alpha$ be an optimal ordering of $G$ and let
$X=\{\alpha^{-1}(1),\alpha^{-1}(2),\ldots,
\alpha^{-1}(\alpha(x)-1)\}.$ Note that for all $ a \in N(x)-X$ we
have $\prf_{\alpha}(G,a) \geq \prf_{\alpha_x}(G-x,a)+1$.
Furthermore, $\prf_{\alpha}(G,x) \geq |N(x) \cap X|$. Thus,


$\begin{array}{rcl} \prf_{\alpha}(G)-\prf_{\alpha_x}(G-x) & \geq &
\prf_{\alpha}(G,x)+\sum\limits_{a\in
N(x)-X}(\prf_{\alpha}(G,a)-\prf_{\alpha_x}(G-x,a)) \\
& \geq & |N(x) \cap X| + |N(x)-X| = d(x) \\
\end{array}$

Hence, $\prf_{\alpha}(G)\ge \prf_{\alpha_x}(G-x) + d(x).$
\end{proof}

\2


Theorem \ref{2_connect} gives a lower bound of the profile of a
2-edge-connected graph, which is important for our FPT algorithm.
Lin and Yuan \cite{linAMAS10} used a concise and elegant argument to
show that $\prf(G)\ge k(2n-k-1)/2$ for every $k$-connected graph $G$
of order $n$. Their argument uses Menger's Theorem in a clever way,
yet the argument cannot be used to prove our bound. Instead of
Menger's Theorem we will apply the following well-known
decomposition of 2-edge-connected graphs (see, e.g., Theorem 4.2.10
in \cite{west2001}) called a {\em closed-ear decomposition}.

\begin{theorem} \label{ear_decomp}
Any $2$-edge-connected graph $G$ has a partition of its edges
$E_1,E_2,\ldots,E_r$, such that $G_i=G[E_1 \cup E_2 \cup \ldots
\cup E_i]$ is $2$-edge-connected for all $i=1,2,3,\ldots,r$.
Furthermore, $E_j$ induces either a path with its endpoints in
$V(G_{j-1})$ but all other vertices in $V(G_j)-V(G_{j-1})$ or a
cycle with one vertex in $V(G_{j-1})$ but all other vertices in
$V(G_j)-V(G_{j-1})$ for every $j=2,3,\ldots,r$. Moreover, $G_1$ is
a cycle and every cycle of $G$ can be $G_1$.
\end{theorem}

\begin{theorem} \label{2_connect}
If $G$ is a $2$-edge-connected graph of order $n$, then $\prf(G)
\geq \frac{3n-3}{2}$.
\end{theorem}
\begin{proof} Let $\alpha$ be an optimal ordering of $V(G)$ and let $y$ be
the vertex with $\alpha(y)=n$. Since $G$ is 2-edge-connected, $y$ is
contained in a cycle $C.$ By Theorem \ref{ear_decomp}, $G$ has an
ear-decomposition $E_1,E_2,\ldots,E_r$ such that $G[E_1]=C.$ Let
$G_i=G[E_1 \cup E_2 \cup \ldots \cup E_i]$, which by Theorem
\ref{ear_decomp} are $2$-edge-connected for all $i=1,2,\ldots,r$. We
will prove this theorem by induction. If $r=1$ then the Theorem
holds by Lemma \ref{prf_cycle} (ii), as $n \geq 3$. So assume that
$r \geq 2$. Let $n_i=|V(G_i)|$ for all $i=1,2,\ldots,r$ and note
that by induction we know that $\prf(G_{r-1}) \geq
\frac{3n_{r-1}-3}{2}$. If $n_r=n_{r-1}$ then $E_r$ is just one edge
and we are done as $\prf(G_{r}) \geq \prf(G_{r-1}) $. So assume that
$a=n_r-n_{r-1}>0$. If $a=1$ and $V(G_r)-V(G_{r-1})=\{x\}$, then by
Lemma \ref{add_vertex} we obtain the following:
$$ \prf(G) \geq \prf(G_{r-1})+d(x) \geq \frac{3n_{r-1}-3}{2} + 2 > \frac{3n-3}{2}. $$
So we may assume that $a \geq 2$. Let $P$ be the path $G_r -
V(G_{r-1})$,
 let  $x$ and $z$ be the endpoints of $P$ such
that $\alpha(x) < \alpha(z)$,  and let $u$ be the neighbor of $x$
in $G_{r-1}$.  Let $j=\min \{ \alpha(q) \mbox{ : } q \in
V(G_{r-1}) \}$, and let $Q=\{p\in V(P):\ \alpha(p)>j\}$ and
$M=\{p\in V(P):\ \alpha(p)<j\}$, which is a partition of $V(P)$.
(Note that $\alpha^{-1}(j) \in V(G_{r-1})$ and recall that
$\alpha^{-1}(n)=y \in V(G_{r-1})$.) Furthermore let $\beta$ denote
the ordering $\alpha$ restricted to $P$ (i.e.,
$\beta=\alpha_{V(G_{r-1})}$) and let $H=G-M$. By  Lemma \ref{G_X}
(with $X=Q$) we obtain the following:
$$\prf_{\alpha_M} (H,V(H)-Q) \geq
\prf_{\alpha_{(M \cup Q)}}(H-Q)+|Q| =
\prf_{\alpha_{V(P)}}(G_{r-1})+|Q| .$$ Now assume that $\alpha(x)<j$
and note that $\prf_{\alpha}(G,u) \geq \prf_{\alpha_M} (H,u) + j-
\alpha(x)$, as $\prf_{\alpha_M} (H,u) \leq \alpha(u)-j$ and
$\prf_{\alpha} (G,u) = \alpha(u)-\alpha(x)$. As $|Q|=|V(P)|-j+1$ and
$\prf_{\alpha} (G,V(P)) \geq \prf_{\beta} (P)$ we obtain the
following:

$\begin{array}{rcl}
 \prf_{\alpha} (G) & = & \prf_{\alpha} (G,V(H)-Q) + \prf_{\alpha} (G,V(P)) \\
  & \geq & \prf_{\alpha_M} (H,V(H)-Q) +  j- \alpha(x) + \prf_{\beta} (P)  \\
  & \geq & \prf_{\alpha_{V(P)}}(G_{r-1})+|Q| + j- \alpha(x) + \prf_{\beta} (P)  \\
  & = & \prf_{\alpha_{V(P)}}(G_{r-1})+|V(P)|-\alpha(x)+1 + \prf_{\beta} (P)  \\
\end{array} $

Now assume that $\alpha(x)>j$. Analogously to the above we get the
following:

$\begin{array}{rcl}
 \prf_{\alpha} (G) & = & \prf_{\alpha} (G,V(H)-Q) + \prf_{\alpha} (G,V(P)) \\
  & \geq & \prf_{\alpha_M} (H,V(H)-Q) + \prf_{\beta} (P)  \\
  & \geq & \prf_{\alpha_{V(P)}}(G_{r-1})+|Q| +  \prf_{\beta} (P)  \\
  & \geq & \prf_{\alpha_{V(P)}}(G_{r-1})+|V(P)|- \alpha(x)+1 + \prf_{\beta} (P)  \\
\end{array} $

So, we always have $ \prf_{\alpha} (G)\geq
\prf_{\alpha_{V(P)}}(G_{r-1})+|V(P)|- \alpha(x)+1 + \prf_{\beta}
(P).$

Add an artificial vertex $u'$ to the end of the ordering $\beta$ and
add the edges $u'x$ and $u'z$. This results in an ordering $\beta'$
of $V(P) \cup \{u'\}$ where $\beta'(u')=|V(P)|+1$. Since we have
created a cycle we note that $\prf_{\beta'}(P \cup u') \ge
2(|V(P)|+1)-3$, by Lemma \ref{prf_cycle} (ii). Since the profile of
$u'$ in $\beta'$ is $|V(P)|+1-\alpha(x)$ we note that
$\prf_{\beta}(P) \ge 2(|V(P)|+1)-3 - (|V(P)|+1-\alpha(x))$. We now
obtain the following:
$$\prf_{\alpha} (G) \geq \prf_{\alpha_{V(P)}}(G_{r-1}) +|V(P)|- \alpha(x)+1
                         +2(|V(P)|+1)-3-(|V(P)|+1-\alpha(x)). $$
By reducing this formula and using the fact that
$\prf_{\alpha_{V(P)}}(G_{r-1}) \geq \frac{3n_{r-1}-3}{2}$, we get
the following:
$$\prf(G) = \prf_{\alpha} (G) \geq \frac{3n_{r-1}-3}{2}+2|V(P)|-1.$$
Since $|V(P)|=a \geq 2$ we note that $2|V(P)|-1 \geq \frac{3a}{2}$,
which implies the desired result. \end{proof}

\begin{theorem} \label{k_seperate}
Let $G=(V,E)$ be a connected graph of order $n$, let $\prf(G) \leq
n-1+k$ and let $\alpha$ be an optimal ordering of $G$. Let
$V_1,V_2,\ldots V_t$ be a partition of $V$ such that $|V_1|,|V_t|\ge
k+2$ and there is only one edge $x_iy_i$ between $G[V_1\cup V_2\cup
\cdots \cup V_i]$ and $G[V_{i+1}\cup V_{i+2}\cup \cdots \cup V_t]$
for each $i=1,2,\ldots t-1.$ Let $x_i\in V_i$ and $y_i\in V_{i+1}$
for each $i=1,2,\ldots t-1$ and let $\alpha^{-1}(1)\in V_1$ or
$\alpha^{-1}(n)\in V_t$. Let an ordering $\alpha'$ of $G$ be defined
as follows: $\alpha'_{V-V_i}=\alpha_{V-V_i}$ for each $i=1,2,\ldots
t$, and $\alpha'(v_i)<\alpha'(v_{i+1})$  for each $i=1,2,\ldots
,t-1$. Then $\alpha'$ is optimal.
\end{theorem}
\begin{proof}  Consider
first the case of $t=2.$ Let $xy=x_1y_1$, $X=V_1,\ Y=V_2$. Let
$\alpha$ be an optimal ordering of $G$ and let $\alpha^{-1}(n)=y'
\in Y$ (the case $\alpha^{-1}(1)\in X$ is treated similarly). Let
$x'$ be the vertex with $\alpha(x')=1$. If $x' \in Y$, then Lemma
\ref{G_X} implies that $\prf_{\alpha}(G,Y) \geq
\prf_{\alpha_X}(G-X,Y) + |X|$. Since $\prf_{\alpha}(G,X)\ge
\prf_{\alpha_Y}(G[X]) \geq |X|-1$ and $\prf_{\alpha_X}(G[Y]) \geq
|Y|-1$ (both by Lemma \ref{prf_cycle} (i)) and $|X| \geq k+2$, we
conclude that $\prf_{\alpha}(G) \geq |X|+|Y|+k$, a contradiction.
Therefore, $x' \in X$.

Let $i=\min\{\alpha(y''):\ y'' \in Y\}$ and let
$j=\max\{\alpha(x''):\ x'' \in X\}$. Assume for the sake of
contradiction that $i<j$. Let $I=\alpha^{-1}(\{i,i+1,\ldots ,j\})$.
Recall that $\alpha'$ is defined as follows: $\alpha'_X=\alpha_X$
and $\alpha'_Y=\alpha_Y$ but $\alpha'(x'')< \alpha'(y'')$ for all
$x'' \in X$ and $y'' \in Y$. We will prove that $\alpha'$ is
optimal.

Let $L=G[X \cup (Y \cap I)]$ and let $G'=L$ if $xy\not\in E(L)$ and
$G'=L-xy$, otherwise. Let $\beta=\alpha_{V(G)-V(G')}$ (so $\beta$ is
equal to $\alpha$, except we have deleted the last $n-j$ vertices in
the ordering). Note that by Lemma \ref{G_X} (used with the set $Y
\cap I$) we get that $\prf_{\beta}(G',V(G')-(Y \cap I)) \geq
\prf_{\beta_{Y \cap I}}(V(G')-(Y \cap I)) +|Y \cap I|$. This implies
the following: $\prf_{\alpha}(G-xy,X) \geq
\prf_{\alpha_{Y}}(G[X])+|Y \cap I|.$

Analogously we obtain that $\prf_{\alpha}(G-xy ,Y) \geq
\prf_{\alpha_{X}}(Y)+|X \cap I|$, which implies the following:
\begin{equation}\label{eq2} \prf_{\alpha}(G-xy) \geq
\prf_{\alpha_{Y}}(X)+\prf_{\alpha_{X}}(Y)+|I| =
\prf_{\alpha'}(G-xy)+(j-i+1) \end{equation} If
$\alpha(x)>\alpha(y)$, then the above implies the following
contradiction, as $\alpha'(y)-\alpha'(x)< j-i+1$.
$$  \prf_{\alpha}(G)  \geq  \prf_{\alpha}(G-xy)
                      \geq  \prf_{\alpha'}(G-xy)+(j-i+1)
                      >  \prf_{\alpha'}(G) . $$
Therefore we may assume that $\alpha(x)<\alpha(y)$.
 Let $l=\min\{\alpha(z):\ z \in N[y]-\{x\}\}$  and let
 $L=\alpha^{-1}(\{\alpha(x),\alpha(x)+1,\alpha(x)+2, \ldots ,l-1\})$.
Note that $L=\emptyset$ if $l<\alpha(x)$. By the definition of $L$
and the inequality in (\ref{eq2}), we get the following:
$$ \prf_{\alpha}(G)  =  \prf_{\alpha}(G-xy) + |L|  \geq
                        \prf_{\alpha'}(G-xy) + |I| + |L| $$
When we add the edge $xy$ to $G-xy$, we observe that, in the
ordering $\alpha'$, the profile of $y$ will increase by one for
every vertex from $Y$ with an $\alpha$-value less then $l$ and every
vertex in $X$ with an $\alpha$-value larger than $\alpha(x)$. This
is exactly the set $R_1 \cup R_2 \cup R_3 \cup R_4$, where

$\begin{array}{rcl} R_1 & = & \{y'' \in Y \mbox{ : } \alpha(y'') <
l  \mbox{ and } \alpha(x) <
\alpha(y'')\} \\
R_2 & = & \{x'' \in X \mbox{ : } \alpha(x) < \alpha(x'') \mbox{
and } \alpha(x'') <
l \} \\
R_3 & = & \{y'' \in Y \mbox{ : } \alpha(y'') < l  \mbox{ and }
\alpha(y'') <
\alpha(x) \} \\
R_4 & = & \{x'' \in X \mbox{ : } \alpha(x) < \alpha(x'') \mbox{
and } l <
\alpha(x'') \} \\
\end{array}$

Since $R_1 \cup R_2 \subseteq L$ and $R_3 \cup R_4 \subseteq I$
(as $\alpha^{-1}(l) \in Y$) we conclude that
$$\prf_{\alpha}(G) \geq
\prf_{\alpha'}(G)+|I|+|L|-|R_1|-|R_2|-|R_3|-|R_4| \geq
\prf_{\alpha'}(G).$$

Now let $t\ge 3$. Let $X=\cup_{i=1}^{t-1}V_i$ and $Y=V_t.$ By the
case $t=2$, the following ordering $\beta$ is optimal:
$\beta_X=\alpha_X$, $\beta_Y=\alpha_Y$, and $\beta(x)<\beta(y)$ for
each $x\in X,y\in Y.$ Now let $X'=\cup_{i=1}^{t-2}V_i,\
Y'=V_{t-1}\cup V_t.$ By the case $t=2$, the following ordering
$\beta'$ is optimal: $\beta'_{X'}=\beta_{X'}$,
$\beta'_{Y'}=\beta_{Y'}$, and $\beta'(x')<\beta'(y')$ for each
$x'\in X',y\in Y'.$ Combining the properties of $\beta$ and
$\beta'$, we obtain that $\beta'_{Y'}=\alpha_{Y'}$,
$\beta'_{V-V_{t-1}}=\alpha_{V-V_{t-1}}$,
$\beta'_{V-V_{t}}=\alpha_{V-V_{t}}$, and
$\beta'(x')<\beta'(v_{t-1})<\beta'(v_{t})$ for each $x'\in
X',v_{t-1}\in V_{t-1},v_t\in V_t.$ Continuation of this argument
allows us to show that $\alpha'$ is an optimal ordering.
 \end{proof}

A {\em bridgeless component} of a graph $G$ is a maximal induced
subgraph of $G$ with no bridges. We call a connected graph $G$ a
\emph{chain of length $t$} if the following holds: (a) $G$ has
bridgeless components $C_i$, $1\leq i \leq t$ such that
  $V(G)=\bigcup_{i=1}^t V(G)$, and
(b) $C_i$ is linked to $C_{i+1}$ by a bridge, $1\leq i \leq t-1$. A
component $C_i$ is {\em nontrivial} if $|V(C_i)|>1$, and {\em
trivial}, otherwise. An ordering $\alpha$ of $G$ is \emph{special}
if for any two vertices $x,y\in V(G)$ and $x\in V(C_i), y\in
V(C_j)$, $i<j$ implies $\alpha(x)<\alpha(y)$.

\begin{lemma}\label{lem:chain}
Let $G$ be a chain of order $n$ and let $\eta$ be the total number
of vertices in the nontrivial bridgeless components of $G$. Let
$\alpha$ be a special ordering of $G$ with $\prf_\alpha(G)\leq
n-1+k$. Then $\eta \leq 3k$.
\end{lemma}
\begin{proof} We show $\eta \leq 3k$ by induction on $n.$ Suppose that
$G$ has a trivial component. If $C_1$ is trivial, then $G-C_1$ is a
chain with $\prf_{\alpha_{V(C_1)}}(G-C_1)\leq n'-1+k$, where
$n'=n-1.$ Thus, by induction hypothesis, $\eta\leq 3k$. Similarly,
we prove $\eta\leq 3k$ when $C_t$ is trivial. Assume that $C_i$,
$1<i<t$, is trivial. Let $C_i$ be adjacent to $x\in V(C_{i-1})$ and
$y\in V(C_{i+1})$. Consider $G'$ obtained from $G$ by deleting $C_i$
and appending edge $xy$. Observe that $G'$ is a chain and
$\prf_{\alpha_{V(C_i)}}(G')\leq n'-1+k$, where $n'=n-1.$ Thus, by
induction hypothesis, $\eta\leq 3k$. So, now we may assume that
$\eta =n.$

Let $C_1,\dots,C_t$ denote the bridgeless components of $G$ as in
the definition above. Let $n_i=\Card{V(C_i)}$. If $t=1$, then by
Lemma~\ref{2_connect} we have $n\leq 2k+1$ and we are done as $k\ge
1$. Now assume $t\geq 2$. Let $G'=G-V(C_t)$ and $n'=n-n_t$.  Observe
that $G'$ is a chain and $\alpha_{V(C_t)}$ is a special ordering of
$G'$. Let $k_t=\prf_\alpha(G,V(C_t))-n_t+1$ and let
$k'=\prf_\alpha(G,V(G'))-n'+1$. We have $k_t+k'-1\leq k$.
Lemma~\ref{2_connect} implies that $$n_t-1+k_t =
\prf_\alpha(G,V(C_t))\geq \prf(C_t) +1 \geq \frac{3n_t-3}{2}
+1=\frac{3n_t-1}{2},$$ and thus $k_t\geq \frac{n_t+1}{2}$ and
$n_t\leq 2k_t
  -1$.  Since $n_t\geq 3$, we have $k_t\geq 2$.
By induction hypothesis, $n'\leq 3k'$.  Thus $n=n'+n_t\leq
3(k-k_t+1) + 2k_t-1 \leq 3k-k_t+2\leq 3k$.
\end{proof}

A connected component of a graph $G$ is called {\em nontrivial} if
it has more than one vertex.
\begin{lemma}\label{lem:nontriv}
  Let $G=(V,E)$ be a connected graph of order $n$, let $X\subseteq V$
  such that $G[X]$ is connected. Let $G_1,\dots,G_r$ denote the
  nontrivial connected components of $G-X$. Assume that $\Card{V(G_i)}\leq
  \Card{V(G_{i+1})}$ for $1\leq i \leq r-1$.
 If $k+n-1\geq \prf(G)$, then $k+2\geq r$ and $2k\geq
  \sum_{i=1}^{r-2}\Card{V(G_i)}$.
\end{lemma}
\begin{proof}
  The result holds vacuously true if $r<3$, hence assume $r\geq 3$.
  Let $\alpha$ be an optimal ordering of $G$.  Let $I=\SB 1\leq i \leq
  r \SM V(G_i) \cap \{\alpha^{-1}(1), \alpha^{-1}(n)\}=\emptyset \SE$.
  Clearly $\Card{I}\geq r-2$.  Let $Y=X\cup \bigcup_{i\notin I}V(G_i)$
  and $Z=V\setminus Y$. Observe that $G[Y]=G-Z$ is connected and
  $G_i$, $i\in I$, are exactly the nontrivial components of $G-Y$.
  Since also $\{\alpha^{-1}(1),\alpha^{-1}(n)\}\subseteq Y$, Lemma~\ref{G_X}
  applies.
Thus we get
 $$
    \prf(G)= \prf_\alpha(G) \ge \prf_\alpha(G,V-Z) +
  \sum_{i\in I}\prf_\alpha(G,V(G_i))
           \geq
          \prf(G-Z)+\Card{Z} + \sum_{i\in I}\prf(G_i).
$$
  Furthermore, by Lemma \ref{prf_cycle} (i),
  \begin{eqnarray*}
     k &\geq& \prf(G)-n+1
     \geq \prf(G-Z)+\Card{Z} -\Card{Y} + (\sum_{i\in
    I}\prf(G_i)-\Card{V(G_i)})+1 \\
           &\geq& (\prf(G[Y]) -\Card{Y}) + \Card{Z} - \Card{I}+1\ge -1+ \Card{Z} - \Card{I}+1.
  \end{eqnarray*}
 Hence
  $k\geq \Card{Z} - \Card{I}$.  However, since the components $G_i$
  are nontrivial, $\Card{Z}\geq 2\Card{I}$. Thus, $\Card{I}\leq k$
  and  $|Z|\le k+|I|\le 2k$.
\end{proof}

\section{Vertices of degree 1}

In this section, $G$ denotes a connected graph of order $n$. For an
ordering $\alpha$ of $G$ let $E_\alpha(G)$ denote the set of edges
$uv$ of $G$ such that $\alpha(u)=\min_{w\in N[v]} \alpha(w)$ and
$u\neq v$. The \emph{length $\ell_\alpha(uv)$ of an edge $uv\in
E(G)$
  relative to $\alpha$} is $|\alpha(u)-\alpha(v)|$ if $uv\in
E_\alpha(G)$, and $0$ if $uv\notin E_\alpha(G)$.  Observe that $
\prf_\alpha(G)=\sum_{e\in E(G)}\ell_\alpha(e).
$

Let $X,Y$ be two disjoint sets of vertices of  $G$ and let $\alpha$
be an ordering of $G$. We say that $(X,Y)$ is an \emph{$\alpha$\hy
consecutive pair} if there exist integers $a,b,c$ with $1\leq a < b
< c \leq n$ so that $X=\SB x\in V(G) \SM a\leq \alpha(x) \leq b-1
\SE$ and $Y=\SB y\in V(G) \SM b\leq \alpha(y) \leq c \SE$. By
$\swap_{Y,X}(\alpha)$ we denote the ordering obtained from $\alpha$
by swapping the $\alpha$\hy consecutive pair $(X,Y)$. For a set
$X\subseteq V(G)$ let $E^r_\alpha(X)$ (respectively,
$E^l_\alpha(X)$) denote the set of edges $uv\in E_\alpha$ with $u\in
X$, $v\in V(G)\setminus X$, and $\alpha(u)<\alpha(v)$ (respectively,
$\alpha(u)>\alpha(v)$).

\begin{lemma}\label{lem:swap2}
  Let $\alpha$ be an ordering of $G$ and $(X,Y)$ an $\alpha$\hy
  consecutive pair such that there are no edges between $X$ and $Y$.
  If $\Card{E^l_\alpha(X)} \leq \Card{E^r_\alpha(X)}$ and
  $\Card{E^l_\alpha(Y)} \geq \Card{E^r_\alpha(Y)}$, then for
  $\beta=\swap_{Y,X}(\alpha)$ we have $\prf_\beta(G)\leq
  \prf_\alpha(G)$.
\end{lemma}
\begin{proof}
  Observe that $E_\alpha(G)=E_\beta(G)$.  Moreover, the only edges of
  $E_\alpha(G)$ that have different length in $\alpha$ and in $\beta$ are
  the edges in $E^l_\alpha(Y) \cup E^r_\alpha(Y) \cup E_\alpha^l(X)
  \cup E_\alpha^r(X)$. Observe that $\ell_\beta(e)=\ell_\alpha(e)+|Y|$,
$\ell_\beta(e')=\ell_\alpha(e')-|Y|,$
$\ell_\beta(f)=\ell_\alpha(f)-|X|$,
$\ell_\beta(f)=\ell_\alpha(f)+|X|$ for each $e\in E^l_\alpha(X)$,
$e'\in E^r_\alpha(X)$, $f\in E^l_\alpha(Y)$ and $f'\in
E^r_\alpha(Y).$ Using these relations and the inequalities
$\Card{E^l_\alpha(X)} \leq
  \Card{E^r_\alpha(X)}$ and $\Card{E^l_\alpha(Y)} \geq
  \Card{E^r_\alpha(Y)}$, we obtain
 $\prf_\beta(G)\leq
  \prf_\alpha(G)$.
\end{proof}

\begin{lemma}\label{lem:swap1}
  Let $\alpha$ be an ordering of $G$ and $(\{x\},Y)$ an $\alpha$\hy
  consecutive pair such that $x$ has a neighbor $z$ of degree $1$
  with $\alpha(z)>\alpha(y)$ for all $y\in Y$.  If
  $\Card{E^l_\alpha(Y)} \geq \Card{E^r_\alpha(Y)}$, then for
  $\beta=\swap_{Y,\{x\}}$ we have $\prf_\beta(G)\leq \prf_\alpha(G)$.
\end{lemma}
\begin{proof}
  If there are no edges between $x$ and vertices in $Y$ then
  the result follows from Lemma~\ref{lem:swap2} since
  $\Card{E^l_\alpha(\{x\})} \leq 1\le \Card{E^r_\alpha(\{x\})}$.

  Now consider the case where $E_\alpha^l(\{x\})=\{wx\}$ for a vertex
  $w$.  It follows that $E_\beta(G)\subseteq E_\alpha(G)$.  Moreover, we
  have
  $\sum_{e\in E^l_\beta(Y)\cup E^r_\beta(Y)}\ell_\beta(e) \leq \sum_{e\in
    E^l_\alpha(Y)\cup E^r_\alpha(Y)}\ell_\alpha(e)$
  and
  $\ell_\beta(wx)+\ell_\beta(xz) \leq
  \ell_\alpha(wx)+\ell_\alpha(xz)$.
  Hence the result also holds true in that case.

  It remains to consider the case where $x$ has neighbors in $Y$ and
  $E_\alpha^l(\{x\})=\emptyset$. Let $y,y'$ be the neighbors of $x$
  in $Y$ with largest $\alpha(y)$ and smallest $\alpha(y')$.  Now
  $E_\beta(G)\setminus E_\alpha(G)=\{xy'\}$, and
  $\ell_\beta(xy')+\ell_\beta(xz) \leq \ell_\alpha(xz)$.  Thus,
  $\prf_\beta(G)\leq \prf_\alpha(G)$.
\end{proof}

\begin{lemma}\label{lem:swap3}
  Let $\alpha$ be an ordering of $G$ and let $(\{x\},Y)$ be an
  $\alpha$\hy consecutive pair.  Let all vertices in $Y$ be
  of degree $1$ and adjacent with $x$.
Then for $\beta=\swap_{Y,\{x\}}(\alpha)$ we have
  $\prf_\beta(G)\leq \prf_\alpha(G)$.
\end{lemma}
\begin{proof}
  Let $y,y'$ denote the vertex in $Y$ with largest $\alpha(y)$ and
  smallest $\alpha(y')$.  Observe $\ell_\alpha(yx)=\Card{Y}$.
First assume that $E^l_\alpha(\{x\})$ contains an edge $zx$.  We
have
  $E_\beta(G)\subseteq E_\alpha(G) \setminus \{xy\}$, and
  $\ell_\beta(e)\leq \ell_\alpha(e)$ holds for all $e\in
  E_\beta(G)\setminus \{xz\}$.
  Since $\ell_\beta(zx)=\ell_\alpha(zx)+\ell_\alpha(xy)$, the result follows.

  Next assume that  $E^l_\alpha(\{x\})=\emptyset$.
  We have
  $E_\beta(G)\subseteq (E_\alpha(G) \setminus \{xy\}) \cup \{xy'\}$, and
  $\ell_\beta(e)\leq \ell_\alpha(e)$ holds for all $e\in
  E_\beta(G)\setminus \{xy'\}$.
  Since $\ell_\beta(xy')=\ell_\alpha(xy)$, the result follows.
 \end{proof}

 For $x\in V(G)$ let $N_1(X)$ denote the set of neighbors of $x$ that
 have degree $1$.  We say that an ordering $\alpha$ of $G$ is
 \emph{conformal for a vertex $x$} of $G$ if $\SB \alpha(w) \SM w\in
 N_1(x) \SE$ forms a (possibly empty) interval and
$\alpha(w)<\alpha(x)$ holds for all $w\in N_1(x)$. We say that
$\alpha$ is \emph{conformal for a graph $G$} if it is
   conformal for all vertices of $G$.

\begin{theorem}\label{the:conformal}
  For every connected graph $G$ there exists an optimal ordering which is conformal.
\end{theorem}
\begin{proof}
  Let $\alpha$ be an optimal ordering of $G$. Let $x$ be a
  vertex of $G$ for which $\alpha$ is not conformal.  We apply the
  following steps to $\alpha$, until we end up with an optimal
  ordering which is conformal for $x$.  In each step we transform
  $\alpha$ into an optimal ordering $\beta$ in such a way that
  whenever $\alpha$ is conformal for a vertex $x'$, so is $\beta$.
  Hence, we can repeat the procedure for all the vertices one after
  the other, and we are finally left with an optimal ordering which is
  conformal.

  Let $w_1,w_2\in N_1(x)\cup \{x\}$ with minimal $\alpha(w_1)$ and
  maximal $\alpha(w_2)$. We call a set $B\subseteq N_1(x)$ a
  \emph{block} if $\SB \alpha(b) \SM b\in B\SE$ is a nonempty interval
  of integers.  A block is \emph{maximal} if it is not properly
  contained in another block.

  \emph{Step 1.}  Assume that there exist $\alpha$\hy consecutive
  pairs $(\{x\},Y)$, $(Y,Z)$ with the following properties:
  (a) $Y$ and $Z$ are nonempty;
  (b) $Y \cap N_1(x)=\emptyset$;
  (c)  $Z$ is a maximal block.
    By assumption, there is a $z\in Z$ such that $xz\in E(G)$ and
  $\alpha(z)>\alpha(y)$ holds for all $y\in Y$.  Moreover, there are
  no edges between $Y$ and $Z$ and $E^r_\alpha(Z)=\emptyset$.  If
  $\Card{E^l_\alpha(Y)} \geq \Card{E^r_\alpha(Y)}$, then we put
  $\beta=\swap_{Y,\{x\}}(\alpha)$, otherwise we put
  $\beta=\swap_{Z,Y}(\alpha)$.  It follows from
  Lemmas~\ref{lem:swap1} and \ref{lem:swap2}, respectively, that
  $\beta$ is optimal.

  \emph{Step 2.} Assume that there exists an $\alpha$\hy consecutive
  pair $(\{x\},Y)$ such that $Y$ is a maximal block.
  We put $\beta=\swap_{Y,\{x\}}(\alpha)$.
  If follows by Lemma~\ref{lem:swap3} that $\beta$ is optimal.

  \emph{Remark.} If neither Step 1 nor Step 2 can be applied, then
  $\alpha(w_2)<\alpha(x)$.

  \emph{Step 3.}  Assume that there exist $\alpha$\hy consecutive
  pairs  $(X,Y)$, $(Y,Z)$ with the following properties:
  (a) $X$ and $Z$ are maximal blocks;
  (b) $Y\subseteq V(G)\setminus  N_1(X)$;
  (c) $w_1\in X$.
Note that there are no edges between $X$ and $Y$ and no edges
  between $Y$ and $Z$.  Furthermore, we have $E^l_\alpha(X)=\emptyset$
  and $E^r_\alpha(Z)=\emptyset$ (the latter follows from Property (c)).
If $\Card{E^l_\alpha(Y)} \geq \Card{E^r_\alpha(Y)}$, then we put
  $\beta=\swap_{Y,X}(\alpha)$, otherwise we put
  $\beta=\swap_{Z,Y}(\alpha)$.  In both cases it follows from
  Lemma~\ref{lem:swap2} that $\beta$ is optimal.

  \emph{Remark.}  If none of the above Steps 1, 2, or 3, applies, then
  $\alpha$ is conformal for $x$.
  \end{proof}

  Note that when applying the procedure of the above proof, it is
  possible that we end up with exactly one maximal block $X$ such that
  for a nonempty set $Y$ the pairs $(X,Y)$ and $(Y,\{x\})$ are
  $\alpha$\hy consecutive.  If $\Card{E^l_\alpha(Y)} <
  \Card{E^r_\alpha(Y)} < \Card{E^r_\alpha(\{x\})}$, then we can
  neither swap $X$ and $Y$ nor $Y$ and $\{x\}$ without increasing the
  cost of the profile.

\section{Kernelization}

For technical reasons, in this section we will deal with a special
kind of weighted graphs, but they will be nothing else but compact
representations of (unweighted) graphs.

We consider a \emph{weighted graph} $G=(V,E,\rho)$ whose vertices
$v$ of degree $1$ have an arbitrary positive integral weight
$\rho(v)$, vertices $u$ of degree greater than one have weight
$\rho(u)=1$. The {\em weight} $\rho(G)$ of $G=(V,E,\rho)$ is the sum
of weights of all vertices of $G$. An {\em ordering} of a weighted
graph $G=(V,E,\rho)$ is an injective mapping $\alpha:V\rightarrow
\{1,\dots,\rho(G)\}$ such that for every vertex $v\in V$ of degree
$1$ we have $\alpha(v)\neq \rho(G)$ and for all $u\in V$ we have
$\alpha(u)\notin \{\alpha(v)+1,\dots,\alpha(v)+\rho(v)-1\}$.  The
{\em profile} $\prf(G)$ of a weighted graph is defined exactly as
the profile of an unweighted graph.

A weighted graph $G=(V,E,\rho)$ corresponds to an unweighted graph
$G^u$, which is obtained from $G$ by replacing each vertex $v$ of
degree 1 ($v$ is adjacent to a vertex $w$) with $\rho(v)$ vertices
adjacent to $w$. By Theorem \ref{the:conformal} and the definitions
above, $\prf(G)=\prf(G^u)$ and an optimal ordering of $G$ can be
effectively transformed into an optimal ordering of $G^u$. Also,
$\rho(G)=|V(G^u)|.$ The correspondence between $G$ and $G^u$ allows
us to use the results given in the previous sections.

\begin{krule}\label{krule:consec}
  Let $G$ be a weighted graph and $x$ a vertex of $G$ with
  $N_1(x)=\{v_1,\dots,v_r\}$, $r\geq 2$.
  We obtain the weighted graph $G_0=(V_0,E_0,\rho_0)$, where
  $G_0=G-\{v_2,\dots,v_r\}$ and $\rho_0(u)=\rho(u)$ for $u\in
  V_0\setminus \{v_1\}$ and $\rho_0(v_1)=\sum_{i=1}^r\nu(v_i)$.
\end{krule}

The next lemma follows from Theorem~\ref{the:conformal}.
\begin{lemma}
  Let $G$ be a weighted connected graph and $G_0$ the weighted graph obtained from
  $G$ by Kernelization Rule~\ref{krule:consec}.
Then $\prf(G)=\prf(G_0)$, and an optimal ordering $\alpha_0$ of
$G_0$ can be effectively transformed into an optimal ordering
$\alpha$ of $G$.
\end{lemma}

Let $e$ be a bridge of a weighted  connected graph $G$ and let
$G_1,G_2$ denote the connected components of $G-e$. We define the
\emph{order} of $e$ as $\min\{\rho(G_1),\rho(G_2)\}$.

Let $v$ be a vertex of a (weighted) graph $G$. We say that $v$ is
\emph{$k$\hy
  suppressible} if the following conditions hold:
(a) $v$ forms a trivial bridgeless component of $G$; (b) $v$ is of
degree $2$ or $3$; (c) there are exactly two bridges $e_1, e_2$ of
order at least $k+2$ incident with $v$; (d) if there is a third edge
$e_3=vw$ incident with $v$, then $w$ is a vertex of degree $1$.

\begin{krule}[w.r.t.\ parameter $k$]\label{krule:supress}
Let $v$ be a $k$\hy suppressible vertex of a weighted graph
$G=(V,E,\rho)$ and let $xv,yv$ be the bridges of order at least
$k+2$. From $G$ we obtain a weighted graph by removing $\{v\}\cup
N_1(v)$ and adding the edge $xy$.
\end{krule}

\begin{lemma}\label{rule2lem}
Let $G=(V,E,\rho)$ be a weighted connected graph with $\prf(G)\leq
\rho(G)-1+k$ and $G'$ the weighted graph obtained from $G$ by means
of Kernelization Rule~\ref{krule:supress} with respect to parameter
$k$. Then $\prf(G)-\rho(G)=\prf(G')-\rho(G')$, and an optimal
ordering $\alpha'$ of $G'$ can be effectively transformed into an
optimal ordering $\alpha$ of $G$.
\end{lemma}
\begin{proof}
Let $v$ be a $k$\hy suppressible vertex of $G^u$ and let $xv,yv$ be
the bridges of order at least $k+2$. We consider the case when
$N_1(v)=\{w_1,\ldots,w_r\}\neq \emptyset$; the proof for the case
when $N_1(v)=\emptyset$ is similar. Let $G^u[X]$ and $G^u[Y]$ denote
the components of $G^u-v$ that contain $x$ and $y$, respectively.
Consider an optimal ordering $\alpha$ of $G^u$ and assume that
$\alpha^{-1}(n)\in Y$. By Theorem \ref{the:conformal}, we may assume
that $\alpha(w_i)<\alpha(v)$ for every $1\le i\le r$. Now by
Theorem~\ref{k_seperate}, we can find an optimal ordering $\alpha'$
of $G^u$ such that $\alpha'(x')<\alpha'(w_i)<\alpha'(v)<\alpha'(y')$
for each $x'\in X,\ y'\in Y$ and $i=1,2,\ldots ,r.$

Now it will be more convenient to argue using the weighted graphs
$G$ and $G'$. Using Kernelization Rule~\ref{krule:consec}, we
transform $\alpha'$ into the corresponding optimal ordering of $G$.
For simplicity we denote the new ordering $\alpha'$ as well. Observe
that
$\prf_{\alpha'_{\{v,w\}}}(G',y)=\prf_{\alpha'}(G,y)+\prf_{\alpha'}(G,v)-1-\rho(w).$
Hence, $\prf(G')-\rho(G')\le \prf_{\alpha'_{\{v,w\}}}(G')-\rho(G')=
\prf(G)-\rho(G)$.

Conversely, let $\alpha'$ be an optimal ordering of $G'$. Since the
bridge $xy$ of $G'$ is of order at least $k+2$, we may assume by
Theorem~\ref{k_seperate} that either for all $x'\in X$ and $y'\in Y$
we have $\alpha'(x')<\alpha'(y')$.  It is straightforward to extend
$\alpha'$ into an ordering $\alpha$ of $G$ such that
$\alpha_{\{v,w\}}=\alpha'$ and
$\prf_\alpha(G)=\prf_{\alpha'}(G')+1+\rho(w)$.  Hence
$\prf(G)-\rho(G)\le \prf_{\alpha}(G)-\rho(G)=
\prf_{\alpha'}(G')-\rho(G')$. Thus,
$\prf(G')-\rho(G')=\prf(G)-\rho(G)$.
\end{proof}


\begin{theorem}\label{maint}
  Let $G=(V,E,\rho)$ be a weighted connected graph with $n=\Card{V}$ and
  $m=\Card{E}$. Let $k$ be a positive integer such that
  $\prf(G)\le \rho(G)-1+ k$.
Either one of the Kernelization Rules \ref{krule:consec} and
  \ref{krule:supress} can be applied with respect to parameter $k$, or
  $n\leq 12k+6$ and $m \leq 13k+5$.
\end{theorem}
\begin{proof} For a weighted graph $G=(V,E,\rho)$ let $G^*$ be an unweighted graph
with $V(G^*)=V$ and $E(G^*)=E.$ Observe that $\prf(G^*)\le \prf(G).$
Thus, in the rest of the proof we consider $G^*$ rather than $G$,
but for the simplicity of notation we use $G$ instead of $G^*$.

Assume that none of the Kernelization Rules \ref{krule:consec} and
\ref{krule:supress} can be applied with respect to parameter $k$. We
will show that the claimed bounds on $n$ and $m$ hold.   By Theorem
\ref{bilth} we have $m \leq
  \prf(G)\leq n-1+k$.  Thus, $n\leq 12k+6$ implies $m \leq 13k+5$.
Therefore, it suffices to prove that $n\leq 12k+6$. If $G$ is
bridgeless, then by Lemma \ref{2_connect}, we have $n-1+k\ge \prf(G)
\geq \frac{3n-3}{2}$ and, thus, $n\le 2k+1$. Hence, we may assume
that $G$ has bridges. Let $C_i$, $i=1,\dots,t$, denote the
bridgeless components of $G$ such that at least one vertex in $C_i$
is incident with a bridge of order at least $k+2$. We put
$X=\bigcup_{i=1}^tV(C_i)$.

Suppose that there is a component $C_i$ incident with three or more
bridges of order at least $k+2$. Then, we may assume that there are
three bridges $e_2,e_3,e_4$ of order at least $k+2$ that connect a
subgraph $F_1$ of $G$ with subgraphs $F_2,F_3,F_4$, respectively,
and $V=\cup_{i=1}^4V(F_i).$ Let $\alpha$ be an optimal ordering  of
$G$. Assume without loss of generality that $\alpha^{-1}(1) \not\in
V(F_2)$ and $\alpha^{-1}(n) \not\in V(F_2)$. Let $X=V(F_2)$ and note
that $G-X$ is connected. Therefore Lemmas \ref{G_X} and
\ref{prf_cycle} (i) imply the following:
 $$\prf(G)=\prf_{\alpha}(G,X)+\prf_{\alpha}(G,V-X) \geq
          |X|-1+(|V|-|X|-1)+|X| \geq n+k,$$
which is a contradiction.

Since $G$ is connected, it follows that $G[X]$ is connected. Thus,
$G[X]$ is a chain and we may assume that $C_i$ and $C_{i+1}$ are
linked by a bridge $b_i$ of
 for each $i=1,2,\ldots,t-1.$ Notice that each $b_i$ is of order at least
 $k+2$ in $G.$

  Let $G_1,\dots,G_r$ be the connected components of $G-X$. Observe
  that each $G_i$ ($1\leq i \leq r$) is linked with exactly one $C_j$
  ($1\leq j \leq t$) with a bridge $e_{ij}$. The bridge $e_{ij}$ must
  be of order less than $k+2$, since otherwise $V(G_i)\cap
  X\neq\emptyset$.  Hence
  (**)
  $\Card{V(G_i)}\leq k+1$
  follows for all $i\in \{1,\dots,r\}$. For each $j$, let $IG(j)$ be
the set of indices $i$ such that $G_i$ is linked to $C_j.$

  Let $N=\SB 1\leq i \leq t \SM \Card{V(C_i)}>1\SE$ and $T=\SB 1\leq i
  \leq t \SM \Card{V(C_i)}=1\SE$, i.e., $C_i$ is nontrivial for $i\in
  N$ and trivial for $i\in T$. For $i\in T_i$ let $x_i$ denote the
  single vertex in $C_i$.  Similarly, let $N'=\SB 1\leq i \leq r \SM
  \Card{V(G_i)}>1\SE$ and $T'=\SB 1\leq i \leq r \SM
  \Card{V(G_i)}=1\SE$.
Let $H_j=G[\cup_{i\in IG(j)}V(G_i)\cup V(C_j)]$ for each
$j=1,2,\ldots ,t.$ By Theorem \ref{k_seperate}, we may assume that
there exists an optimal ordering $\beta$ such that
$\beta(h_i)<\beta(h_j)$ for all $i<j$, $h_i\in V(H_i),\ h_j\in
V(H_j).$ Let $\gamma=\beta_{V(G)-X}.$ Clearly, $\gamma$ is a special
ordering of the chain $G[X]$, i.e., $\gamma(c_i)<\gamma(c_j)$ for
all $i<j$, $c_i\in V(C_i),\ c_j\in V(C_j).$

If $G_i$ is nontrivial, then it has a vertex $z$ such that $G_i-z$
is connected and $z$ is not incident to the bridge between $G_i$ and
$G[X]$. If $G_i$ is trivial, let $z=V(G_i)$. In both cases, by Lemma
\ref{add_vertex}, $\prf_{\beta_z}(G-z)\le (n-1)-1+k.$ Repeating this
argument, we conclude that $\prf_{\gamma}(G[X])\le |X|-1+k.$ Now by
Lemma~\ref{lem:chain}, $\sum_{i\in N}\Card{V(C_i)}\leq 3k$.
Lemma~\ref{lem:nontriv} yields that $\Card{N'}\leq k+2.$ Observe
that for each $i\in T$, $x_i$ is linked by a bridge $x_iy_{\pi(i)}$
to at least one nontrivial $G_{\pi(i)}$, where $\pi(i)\neq \pi(i')$
whenever $i\neq i'$. Hence, $\Card{T}\leq k+2$. Thus, $\Card{X}=
\sum_{i\in N}\Card{V(C_i)} + \Card{T} \leq 3k +(k+2)=4k+2$. Using
(**) and Lemma \ref{lem:nontriv}, we have that $\sum_{i\in N'}
\Card{V(G_i)} \leq 2(k+1) + 2k=4k+2$.

Let $Y=\cup_{i=1}^rV(G_i)$.  Since Kernelization
  Rule \ref{krule:consec} cannot be applied, every vertex in $X$ is
  adjacent with at most one $G_i$ with $i\in T'$.  Hence
  $\Card{T'}\leq \Card{X}\leq 4k+2$.  Consequently $\Card{Y}\leq
  2(4k+2) = 8k+4$.  Hence $n=\Card{X}+\Card{Y}\leq 4k+2 + 8k+4
  =12k+6$ follows.
\end{proof}

\begin{corollary}
The problem PAVGV is fixed-parameter tractable.
\end{corollary}

\begin{remark} We see that PAVGV can be solved in time
$O(|V|^2+f(k)),$ where $f(k)=(12+6)!$. It would be interesting to
significantly decrease $f(k),$ but even as it is now our algorithm
is of practical interest because the kernel produced by the two
kernelization rules can be solved using fast heuristics.
\end{remark}

\section{NP-completeness}\label{npsec}


Serna and Thilikos \cite{sernaEATCSB86} asked whether the following
problem is FPT.
 \vspace{-.3cm}
\begin{quote}
  \noindent{\bfseries Vertex Average Profile} (VAP)\\
  \emph{Instance:} A graph $G=(V,E)$.
  \emph{Parameter:} A positive integer $k$.\\
  \emph{Question:} Does $G$ have an ordering of profile  $\le k|V|$?
\end{quote}
 \vspace{-.3cm}
 The following result was announced in \cite{gutinTCS}
without a proof. It implies that VAP is not FPT unless {\rm P}=\NP.

\begin{theorem} Let $k \geq 2$ be a fixed integer. Then it is NP-complete to decide
whether $\prf(H) \leq k|V(H)|$ for a graph $H$.
\end{theorem}
\pf   Let $G$ be a graph and let $r$ be an integer. We know that it
is NP-complete to decide whether $\prf(G) \leq r$. Let $n=|V(G)|$.
Let $k$ be a fixed integer, $k\ge 2$. Define $G'$ as follows: $G'$
contains $k$ copies of $G$, $j$ isolated vertices and a clique with
$i$ vertices (all of these subgraphs of $G'$ are vertex disjoint).
We have $n'=|V(G')|=kn+i + j.$ Observe that prf$(K_i)={i \choose
2}$. By the definition of $G'$, $k\cdot \prf(G)=\prf(G')-\prf(K_i)
                         =\prf(G')-{i \choose 2}.$
Therefore, $\prf(G) \leq r$ if and only if $\prf(G') \leq kr+{i
\choose 2}.$ If there is a positive integer $i$ such that $kr+{i
\choose 2}=kn'$ and the number of vertices in $G'$ is bounded from
above by a polynomial in $n$, then $G'$ provides a reduction from
 to VAP with the fixed $k.$ Observe that $kr+{i \choose 2}\ge
k(kn+i)$ for  $i=2kn$. Thus, by setting $i=2kn$ and $j=r+{1 \over
k} {i \choose 2}-kn- i$, we ensure that $G'$ exists and the number
of vertices in $G'$ is bounded from above by a polynomial in $n$.
\qed

\end{document}